\documentclass[12pt]{article}
\usepackage{amssymb,amsmath,amsthm,amsfonts,amscd}
 \usepackage{graphicx}
\newcommand\be{\begin{equation}}
\newcommand\ee{\end{equation}}
\newcommand\bea{\begin{eqnarray}}
\newcommand\eea{\end{eqnarray}}

\newcommand{\fatalpha}{{\bf \alpha \kern -0.44em \alpha}}
\newcommand{\fatsigma}{{\bf \sigma \kern -0.54em \sigma}}
\newcommand{\tpchi}{{\bf D \kern -0.35em D}}
\newcommand{\llambda}{{\bf \lambda \kern -0.45em \lambda}}

\renewcommand{\theequation}{\arabic{equation}}
\renewcommand{\theequation}{\thesection.\arabic{equation}}
\bibliography{plain}
\title{\bf Two-qutrit Entanglement Witnesses and Gell-Mann Matrices}\vspace{20mm}
\author{ M. A. Jafarizadeh $^{a,b,c}$
 \thanks{E-mail:jafarizadeh@tabrizu.ac.ir}  ,
 Y. Akbari  $^{a,b}$
 \thanks{E-mail:y-akbari@tabrizu.ac.ir} ,
  N. Behzadi  $^{a,b}$
 \thanks{E-mail:behzadi@tabrizu.ac.ir},
\\ $^a${\small Department of Theoretical Physics and Astrophysics,
Tabriz University, Tabriz 51664, Iran.} \\ $^b${\small Institute
for Studies in Theoretical Physics and Mathematics, Tehran
19395-1795, Iran.} \\ $^c${\small Research Institute for
Fundamental Sciences, Tabriz 51664, Iran. }} \pagebreak

\vspace{20mm}
\begin{document}
\maketitle \vspace{15mm}
\newpage
\begin{abstract}
The Gell-Mann $\lambda$ matrices for Lie algebra su(3) are  the
natural basis for the Hilbert space of Hermitian operators acting
on the states of a three-level system(qutrit).  So the
construction of EWs for two-qutrit states by using these matrices
may be an interesting problem. In this paper, several two-qutrit
EWs are constructed based on the Gell-Mann matrices by using the
linear programming (LP) method exactly or approximately. The
decomposability and non-decomposability of constructed EWs are
also discussed and it is shown that the $\lambda$-diagonal EWs
presented in this paper are all decomposable but by adding
$\lambda$-non-diagonal terms, one can obtain various
non-decomposable EWs.

\end{abstract}


{\bf PACs Index: 03.65.Ud}

\vspace{70mm}
\newpage
\section{Introduction}
Entanglement is one of the interesting features of quantum
systems. It is used as a physical resource in realization of many
quantum information and quantum computation processes such as
quantum cryptography, teleportation, dense coding and quantum key
distribution \cite{nielsen1,ekert1,preskill} which cannot be
realized in classical physics. So a strong motivation has been
raised for the study of entanglement detection in an operational
way. Though the fields of quantum information and quantum
computation have been mainly built on the concept of qubits,
however the exploration of qutrits has been attracted much
attention in the recent years
\cite{narnhofer1,narnhofer2,bertlmann1}. A qutrit, the simplest
generalization of a qubit, is a system whose operators act on a
three-dimensional Hilbert space. Qutrits have some interesting
properties that account for their usefulness: they have improved
the efficiency and security of many quantum information protocols,
and it is expected that systems of entangled qutrits largely
violate the non-locality via Bell tests \cite{klimovl}. Therefore,
the detection of entanglement, that is, distinguishing between
separable and entangled states, for two-qutrit systems is an
important problem.
\par
Among the different criteria to analyze the separability  of
quantum states,  the entanglement witnesses (EWs) are of special
interest since it has been proved that for any entangled state
there exists at least one EW which detects its entanglement
\cite{woron1,horod2}. The EWs are Hermitian operators which have
non-negative expectation values over all separable states and
detect some entangled states. Therefore, for construction of EWs,
one needs to determine the minimum of this expectation value and
to demand its non-negativity which lies in realm of the convex
optimization problem. The linear programming (LP), as a special
case of convex optimization, is one of the very useful approaches
for constructing EWs \cite{doherty1,doherty2,jafar1,jafar2,jafar3,
jafar4,jafar5} since it can be solved by using very efficient
algorithms such as the simplex and interior-point methods ( see
e.g. \cite{boyd,chong}).
\par
Inside the several problems concerning the EWs, the problem of how
to construct EWs by a given set of operators has a great
importance. In this paper, we use the Gell-Mann $\lambda$ matrices
, a basis for the Lie algebra su(3) \cite{pfeifer}, to construct
EWs for two-qutrit states. Two  $\lambda$-diagonal and
$\lambda$-non-diagonal cases are considered and it is shown that
the problem of constructing such EWs can be reduced to a LP
problem. We show that, as far as related to the LP method, the
presented $\lambda$-diagonal EWs are all decomposable while a
large number of $\lambda$-non-diagonal EWs are non-decomposable.
To show the non-decomposability of these EWs, we find a class of
bound entangled states or used known bound entangled states that
can be detected by them. Also, it is noted that in determining the
decomposability or non-decomposability of a given EW, the pure
product states which have zero expectation values with it play a
crucial role.
\par
Determination of the feasible region is the main goal of any
separability problem. So in the method of the present paper, it is
tried to determine the feasible region exactly or approximately.
When the feasible region was determined, any hyper-plane tangent
to it would correspond to an optimal EW. In this way, we get all
possible optimal EWs, including the pre-existent ones obtained by
other methods.
\par
The paper is organized as follows: In Section 2, we review the
basic notions and definitions of EWs relevant to our study and
describe a general approach of constructing EWs by the LP method.
In Section 3, we consider the construction of $\lambda$-diagonal
EWs and Section 4 is devoted to the construction of various
$\lambda$-non-diagonal EWs. The paper is ended with a brief
conclusion and three appendices.

\section{EWs and LP method}
\subsection{Entanglement witnesses}
First let us recall the definition of entanglement and
separability. A bipartite quantum mixed state $\rho\in
{\cal{B}}({\cal{H}})$ (the Hilbert space of bounded operators
acting on $\mathcal{H}={\cal{H}}_{d_{1}}\otimes{\cal{H}}_{d_{2}})$
is called separable if it can be written as a convex combination
of pure product states, that is
\begin{equation}\label{fullsep}
    \rho=\sum_{i} p_{i} | \alpha_{i}^{(1)} \rangle \langle \alpha_{i}^{(1)} |\otimes
    | \alpha_{i}^{(2)} \rangle \langle \alpha_{i}^{(2)}|
    \end{equation}
where $|\alpha_{i}^{(j)}\rangle$ are  arbitrary but normalized
vectors lying in the  $\mathcal{H}_{d_{j}}$, and $p_{i}\geq0$ with
$\sum_{i}p_{i}=1$. When this is not the case, $\rho$ is called
entangled.
 \par
An entanglement witness $W$ is a Hermitian operator such that
$Tr(W\rho_{s})\geq 0 $ for all separable states $\rho_{s}$ and
there exists at least one entangled state $\rho_{e}$ which can be
detected by $W$, that is $ Tr(W\rho_{e})<0 $. The existence of an
EW for any entangled state is a direct consequence of Hahn-Banach
theorem \cite{rudin1} and the fact that the  space of separable
density operators is convex and closed.
\par
Based on the notion of partial transpose, the EWs are classified
into two classes: decomposable (d-EW) and non-decomposable
(nd-EW). An EW $W$ is called decomposable if there exist positive
operators $\mathcal{P}, \mathcal{Q}$ such that
\begin{equation}\label{dew}
 W=\mathcal{P}+\mathcal{Q}^{T_{1}}
\end{equation}
where $T_{1}$ denotes the partial transpose with respect to
partite $1$ and it is non-decomposable if it can not be written in
this form \cite{doherty1}. Clearly a d-EW can not detect bound
entangled states (entangled states with positive partial transpose
(PPT) ) whereas there are some bound entangled states which can be
detected by an nd-EW.
\par
Usually one is interested in finding EWs $W$ which detect
entangled states in an optimal way in the sense that when we
subtract any positive operator from $W$ then it does not remain an
EW anymore \cite{lewen1}. In other words, if there exist
$\epsilon> 0$ and a positive operator $ \mathcal{P}$ such that
$W'=W-\epsilon \mathcal{P}$ is again an EW then we conclude that
$W$ is not optimal, otherwise it is.
\subsection{Linear programming method}
This subsection is devoted to  describing linear programming (LP)
and a general approach  for constructing EWs by the LP method
exactly or approximately \cite{boyd}.
\par
Let us consider a non-positive Hermitian operator of the form
\begin{equation}\label{wgen}
W=a_{_{0}}I+\sum_{i} a_{i}Q_{_{i}}
\end{equation}
where $Q_{_{i}}$ are Hermitian operators and $a_{_{i}}$'s are real
parameters with $a_{_{0}}>0$. In this work, the $Q_{_{i}}$ will be
considered as tesor products of Gell-Mann matrices for Lie algebra
su(3). We will attempt to choose the real parameters $a_{i}$ such
that $W$ becomes an EW. To this aim, we introduce the maps
\begin{equation}\label{varp}
    P_{_{i}}=Tr(Q_{_{i}}\rho_{s})
\end{equation}
for any separable state $\rho_{s}$. The maps $P_{_{i}}$ map the
convex set of separable states into a bounded convex region which
will be named feasible region. The first property of an EW is that
its expectation value over any separable state is non-negative,
i.e., the condition
$$
F_{W}:=Tr(W\rho_{s})=a_{_{0}}+
 \sum_{i}a_{_{i}}P_{_{i}}\geq0
$$
is satisfied for any point $(P_{_{1}},P_{_{2}},...)$ of the
feasible region. In order to satisfy this condition, it is
sufficient that the minimum value of $F_{W}$ be non-negative. We
note that the quantity $F_{W}$ achieves its minimum value for pure
product states, since every separable mixed state $\rho_{s}$ can
be written as a convex combination of pure product states, say
$\rho_{s}=\sum_{i}p_{i} |\gamma_{i}\rangle\langle\gamma_{i}|$ with
$p_{i}\geq0$ and $\sum_{i}p_{i}=1$, whence we have
\begin{equation}
Tr(W\rho_{s})=\sum_{i}p_{i}Tr(W|\gamma_{i}\rangle\langle\gamma_{i}|)\geq
C_{min},
\end{equation}
$$
\mathrm{with}  \quad\quad C_{min}=\min_{_{|\gamma\rangle\in
D_{prod}}} \;Tr(W |\gamma\rangle\langle\gamma|)
$$
where $D_{prod}$ denotes the set of pure product states. Here in
this work we are interested in the EWs that their feasible regions
are of simplex (or at most convex polygon) types. The manipulation
of these EWs  amounts to
\begin{equation}\label{lp}
\begin{array}{c}
\mathrm{minimize}\quad F_{{W}}=a_{_{0}}+
\sum_{i}a_{_{i}}P_{_{i}}\\
\mathrm{subject\; to}\quad
\sum_{i}(c_{_{ij}}P_{_{i}}-d_{_{j}})\geq 0 \quad  j=1,2,...\\
\end{array}
\end{equation}
where $c_{_{ij}}$ and $ d_{_{j}}$  are parameters of hyper-planes
surrounding  the feasible regions. So the problem reduces to a LP
problem.  On the basis of the LP method, minimum of an objective
function $F_{W}$ always  occurs at the vertices of bounded
feasible region. Therefore the vertices of the feasible region
come from pure product states.
\par
It is necessary to distinguish between two cases: (\textbf{a})
exactly soluble, and (\textbf{b}) approximately soluble EWs. In
the case \textbf{a} the boundaries (constraints on $P_{_{i}}$)
come from the finite set of vertices arising from pure product
states and construct a convex polygon, while in the case
\textbf{b} the boundaries may not be hyper-planes. In this case we
approximate the boundaries with hyper-planes and clearly some
vertices of the approximated feasible region do not arise from
pure product states. Both cases can be solved by the well-known
simplex method. The simplex algorithm is a common algorithm used
to solve an optimization problem with a polytope feasible region,
such as a linear programming problem. Here, considering the scope
of this paper, a complete treatment of the simplex algorithm is
unnecessary; for a more complete treatment please refer to any LP
text such as \cite{boyd,chong}.
\section{$\lambda$-diagonal EWs}
\subsection{Exactly soluble EWs}\label{sec1}
In this subsection we consider $\lambda$-diagonal EWs which can be
solved exactly by the LP method. Let us consider a situation that
the Hermitian operator is as follows
\begin{equation}\label{ew}
    W=a_{_{0}}I_{_{3}}\otimes I_{_{3}}+\sum_{k=1}^8
a_{_{k}}\lambda_{_{k}}\otimes\lambda_{_{k}}
\end{equation}
where $\lambda_{_{k}}$'s are the Gell-Mann matrices for Lie
algebra su(3). Evidently when all of its nine eigenvalues are
positive the above operator is positive; otherwise it may be an
EW.
\par
Now we attempt to choose the real parameters $a_{i}$ such that $W$
becomes an EW. To this aim, we need to minimize the expectation
value of $W$ over all separable states and to demand its
non-negativity. To reduce the problem to a LP problem and to
determine the feasible region, we require to know the vertices,
namely the extreme points, of the feasible region. Vertices of the
feasible region come from pure product states and are listed in
the table 1
\begin{table}[h]
\renewcommand{\arraystretch}{1}
\addtolength{\arraycolsep}{-2pt}
$$
\begin{array}{cccccccc|c}\hline
  P_{1} & P_{2} & P_{3} & P_{4} & P_{5} & P_{6} & P_{7} & P_{8} & \mathrm{Product \ state}
  \\ \hline
  \pm1 & 0 & 0 & 0 & 0 & 0 & 0 & \frac{1}{3} & |\lambda_{_{1}};+1\rangle\otimes|\lambda_{_{1}};\pm1\rangle\\
  0 & \pm1 & 0 & 0 & 0 & 0 & 0 & \frac{1}{3} & |\lambda_{_{2}};+1\rangle\otimes|\lambda_{_{2}};\pm1\rangle \\
  0 & 0 & \pm1 & 0 & 0 & 0 & 0 & \frac{1}{3} & |\lambda_{_{3}};+1\rangle\otimes|\lambda_{_{3}};\pm1\rangle \\
  0 & 0 & \frac{1}{4} & \pm1 & 0 & 0 & 0 & \frac{1}{12} & |\lambda_{_{4}};+1\rangle\otimes|\lambda_{_{4}};\pm1\rangle \\
  0 & 0 & \frac{1}{4} & 0 & \pm1 & 0 & 0 & \frac{1}{12} & |\lambda_{_{5}};+1\rangle\otimes|\lambda_{_{5}};\pm1\rangle \\
  0 & 0 & \frac{1}{4} & 0 & 0 & \pm1 & 0 & \frac{1}{12} & |\lambda_{_{6}};+1\rangle\otimes|\lambda_{_{6}};\pm1\rangle \\
  0 & 0 & \frac{1}{4} & 0 & 0 & 0 & \pm1 & \frac{1}{12} & |\lambda_{_{7}};+1\rangle\otimes|\lambda_{_{7}};\pm1\rangle \\
  0 & 0 & 0 & 0 & 0 & 0 & 0 & \frac{4}{3} & |\lambda_{_{8}};\frac{-2}{\sqrt{3}}\rangle\otimes|\lambda_{_{8}};\frac{-2}{\sqrt{3}}\rangle \\
  0 & 0 & 0 & 0 & 0 & 0 & 0 & -\frac{2}{3} & |\lambda_{_{8}};\frac{1}{\sqrt{3}}\rangle\otimes|\lambda_{_{8}};\frac{-2}{\sqrt{3}}\rangle
  \\ \hline
\end{array}
$$
\caption{The product vectors and coordinates of vertices for W.}
\renewcommand{\arraystretch}{1}
\addtolength{\arraycolsep}{-3pt}
\end{table}
where $|\lambda_{_{k}};m\rangle$ is the eigenvector of
$\lambda_{_{k}}$ with eigenvalue $m$ and
$P_{_{k}}=Tr(\lambda_{_{k}}\otimes\lambda_{_{k}}|\gamma\rangle\langle\gamma|)$
in which $|\gamma\rangle$ is an arbitrary pure product state.
\par
The hyper-planes passing through the two vertices
$(0,0,1,0,0,0,0,\frac{1}{3})$ and $(0,0,0,0,0,0,0,\frac{4}{3})$
together with other vertices except the vertex
$(0,0,0,0,0,0,0,-\frac{2}{3})$ have the form
\begin{equation}\label{halfspaces}
\frac{3}{4}\sum_{j=1}^{8}(-1)^{i_{_{j}}}P_{_{j}}-1=0\quad;\quad
i_{_{3}}=i_{_{8}}=0\quad \mathrm{and}\quad i_{_{j}}\in\{0,1\}\quad
\mathrm{for}\quad j\neq 3,8.
\end{equation}
To locate the position of the feasible region with respect to
these hyper-planes, we note that the origin
$(P_{_{1}},...,P_{_{8}})=(0,...,0)$ come from the pure product
state
$$
\begin{array}{c}
  \frac{1}{\sqrt{3}}\left(%
\begin{array}{c}
  1 \\
  1 \\
  1 \\
\end{array}%
\right)\otimes\left(%
\begin{array}{c}
  0 \\
  0 \\
  1 \\
\end{array}%
\right) \\
\end{array}
$$
and hence lies within the feasible region. On the other hand, for
the origin we have
$$
\frac{3}{4}\sum_{j=1}^{8}(-1)^{i_{_{j}}}P_{_{j}}-1=-1<0.
$$
Hence, the feasible region lies in one side of the hyper-planes
(\ref{halfspaces}). To make sure, it is proved that there exists
no pure product state on the other side of these hyper-planes (for
a proof, see appendix II). So the hyper-planes (\ref{halfspaces})
are boundaries of the feasible region.
\par
Each hyper-plane (\ref{halfspaces}) corresponds to a positive
operator or an EW. To see this, note that each hyper-plane
(\ref{halfspaces}) can be rewritten as
\begin{equation}\label{operator}
    Tr([I_{_{3}}\otimes I_{_{3}}-\frac{3}{4}\sum_{_{j=1}}^{8}(-1)^{i_{_{j}}}
    \lambda_{_{j}}\otimes\lambda_{_{j}}]|\gamma\rangle\langle\gamma|)=0
\end{equation}
with $|\gamma\rangle$ a pure product state that corresponds to a
vertex of the feasible region. Hence, the operator
\begin{equation}\label{bwitness}
    I_{_{3}}\otimes I_{_{3}}-\frac{3}{4}\sum_{_{j=1}}^{8}(-1)^{i_{_{j}}}
    \lambda_{_{j}}\otimes\lambda_{_{j}}
\end{equation}
has positive expectation value over any pure product state and we
conclude that it is a positive operator or an EW. To simplify the
analysis, we consider the following cases with this in mind that
in all cases $i_{_{3}}=i_{_{8}}=0$.\\
\textbf{a}- $i_{_{1}}=i_{_{2}}\quad,\quad
i_{_{4}}=i_{_{5}}\quad,\quad i_{_{6}}=i_{_{7}}$.\\
This case contains eight operators which are all positive.\\
\textbf{b}- $i_{_{1}}\neq i_{_{2}}\quad,\quad
i_{_{4}}\neq i_{_{5}}\quad,\quad i_{_{6}}\neq i_{_{7}}$.\\
The eight operators of this case are all d-EWs since they have
some negative eigenvalues and their partial transposes with
respect to the first partite which is equivalent to
$\lambda_{_{i}}\rightarrow -\lambda_{_{i}}$ for $i=2,5,7$, yield
one of the eight positive operators of case \textbf{a}.\\
\textbf{c}-
 $i_{_{1}}= i_{_{2}}\;,\; i_{_{4}}\neq
i_{_{5}}\;,\; i_{_{6}}\neq i_{_{7}}$\quad;\quad $i_{_{1}}\neq
i_{_{2}}\;,\; i_{_{4}}= i_{_{5}}\;,\; i_{_{6}}\neq
i_{_{7}}$\quad;\quad $i_{_{1}}\neq
i_{_{2}}\;,\; i_{_{4}}\neq i_{_{5}}\;,\; i_{_{6}}=i_{_{7}}$.\\
This case contains $24$ operators which are all d-EWs. For
instance, consider the following one
\begin{equation}\label{ewcasec}
    I_{_{3}}\otimes I_{_{3}}-\frac{3}{4}(
    \lambda_{_{1}}\otimes\lambda_{_{1}}+\lambda_{_{2}}\otimes\lambda_{_{2}}
    +\lambda_{_{3}}\otimes\lambda_{_{3}}+\lambda_{_{4}}\otimes\lambda_{_{4}}
    -\lambda_{_{5}}\otimes\lambda_{_{5}}+\lambda_{_{6}}\otimes\lambda_{_{6}}
    -\lambda_{_{7}}\otimes\lambda_{_{7}}+\lambda_{_{8}}\otimes\lambda_{_{8}}).
\end{equation}
It can be written as
$$
\mathcal{P}+\mathcal{Q}^{T_{1}}
$$
in which the positive operators $\mathcal{P}$ and $\mathcal{Q}$
are
$$
\mathcal{P}=3|\psi_{_{1}}\rangle\langle\psi_{_{1}}|\quad,
\quad\mathcal{Q}=3(|\psi_{_{2}}\rangle\langle\psi_{_{2}}|
+|\psi_{_{3}}\rangle\langle\psi_{_{3}}|).
$$
Here $|\psi_{_{i}}\rangle$'s are states that to be defined in
relation (\ref{psi}) of appendix III.\\
\textbf{d}-
 $i_{_{1}}\neq i_{_{2}}\;,\; i_{_{4}}=
i_{_{5}}\;,\; i_{_{6}}=i_{_{7}}$\quad;\quad $i_{_{1}}=
i_{_{2}}\;,\; i_{_{4}}\neq i_{_{5}}\;,\; i_{_{6}}=
i_{_{7}}$\quad;\quad $i_{_{1}}=
i_{_{2}}\;,\; i_{_{4}}=i_{_{5}}\;,\; i_{_{6}}\neq i_{_{7}}$.\\
The $24$ operators of this case are all d-EWs since their partial
transposes are equal to one of the $24$ d-EWs of case \textbf{c}.
\par
In summary, it was shown that among the $64$ boundary hyper-planes
(\ref{halfspaces}) of the feasible region, eight of them
correspond to positive operators and the others correspond to
d-EWs. These $64$ boundaries are hyper-planes by itself while as
it is shown in the sequel, the remaining ones are not hyper-planes
by itself and we have to approximate them by hyper-planes in order
to apply the LP method.
\par
The EW introduced in Eq. (74) of \cite{bertlmann1} for the
isotropic qutrit states and also the EWs in Eqs. (118) and (134)
of \cite{bertlmann2} are among the EWs of case \textbf{b}.
\subsection{Approximately soluble EWs}\label{sec2}
 In this subsection we consider other boundaries of
 the feasible region for EWs of the form (\ref{ew}). We show that these are not
 hyper-planes by itself and we have to approximate them with
 hyper-planes.
 \par
At first, we choose two vertices $(0,0,1,0,0,0,0,\frac{1}{3})$ and
$(0,0,0,0,0,0,0,-\frac{2}{3})$ together with other vertices except
the vertex $(0,0,0,0,0,0,0,\frac{4}{3})$. The hyper-planes passing
through these vertices have the form
\begin{equation}\label{halfspaces1}
\frac{3}{2}\left[(-1)^{i_{_{1}}}P_{_{1}}+(-1)^{i_{_{2}}}P_{_{2}}+P_{_{3}}-P_{_{8}}\right]
+\frac{3}{4}\left[(-1)^{i_{_{4}}}P_{_{4}}+(-1)^{i_{_{5}}}P_{_{5}}+(-1)^{i_{_{6}}}P_{_{6}}
+(-1)^{i_{_{7}}}P_{_{7}}\right]-1=0
\end{equation}
where $i_{_{1}}, i_{_{2}}, i_{_{4}},...,i_{_{7}}\in\{0,1\}$. In
appendix II, we find pure product states that maximizes the left
hand side of (\ref{halfspaces1}) and show that this maximum is not
zero but it is $\frac{3}{8}$. This implies that the hyper-planes
(\ref{halfspaces1}) lie inside the feasible region and so are not
really the boundaries of the feasible region. At first
approximation, we approximate the corresponding real boundaries by
hyper-planes that are parallel to the hyper-planes
(\ref{halfspaces1}) and are tangent to real boundaries of the
feasible region at pure product states giving the maximum value
$\frac{3}{8}$. The consequence of the above approximation is that
some vertices of the approximated feasible region do not come from
pure product states. The EWs corresponding to these approximated
boundary hyper-planes are as follows
\begin{equation}\label{bwitnessapprox}
\begin{array}{c}
  W_{1}=\frac{11}{8}I_{_{3}}\otimes I_{_{3}}-\frac{3}{2}[(-1)^{i_{_{1}}}
    \lambda_{_{1}}\otimes\lambda_{_{1}}+(-1)^{i_{_{2}}}\lambda_{_{2}}\otimes\lambda_{_{2}}
    +\lambda_{_{3}}\otimes\lambda_{_{3}}-\lambda_{_{8}}\otimes\lambda_{_{8}}] \\
  \hspace{1.2cm}-\frac{3}{4}[(-1)^{i_{_{4}}}\lambda_{_{4}}\otimes\lambda_{_{4}}+(-1)^{i_{_{5}}}\lambda_{_{5}}\otimes\lambda_{_{5}}
    +(-1)^{i_{_{6}}}\lambda_{_{6}}\otimes\lambda_{_{6}}+(-1)^{i_{_{7}}}\lambda_{_{7}}\otimes\lambda_{_{7}}].\\
\end{array}
\end{equation}
Among them we study the following one in detail
\begin{equation}\label{bwitnessapprox1}
 W'_{1}=\frac{11}{8}I_{_{3}}\otimes I_{_{3}}-\frac{3}{2}[
 \lambda_{_{1}}\otimes\lambda_{_{1}}+\lambda_{_{2}}\otimes\lambda_{_{2}}
 +\lambda_{_{3}}\otimes\lambda_{_{3}}-\lambda_{_{8}}\otimes\lambda_{_{8}}]
 -\frac{3}{4}[\lambda_{_{4}}\otimes\lambda_{_{4}}+\lambda_{_{5}}\otimes\lambda_{_{5}}
 +\lambda_{_{6}}\otimes\lambda_{_{6}}+\lambda_{_{7}}\otimes\lambda_{_{7}}].\\
\end{equation}
To show the decomposability of $W'_{1}$, we assume that it is
decomposable and write
\begin{equation}\label{witapp}
    W'_{1}=\mathcal{P}+\mathcal{Q}^{T_{_{1}}}.
\end{equation}
Also, we note that $W'_{1}$ vanishes on the pure product states
$|\gamma\rangle=|\alpha\rangle\otimes|\beta\rangle$ in relation
(\ref{pure product}) of appendix II, i.e.,
$$
\begin{array}{c}
  Tr(W'_{1}|\gamma\rangle\langle\gamma|)=\langle\gamma|W'_{1}|\gamma\rangle
=\langle\gamma|\mathcal{P}|\gamma\rangle+\langle\gamma|\mathcal{Q}^{T_{_{1}}}|\gamma\rangle \\
  \hspace{1.3cm}=\langle\gamma|\mathcal{P}|\gamma\rangle
  +\langle\alpha^{*}|\langle\beta|\mathcal{Q}|\alpha^{*}\rangle|\beta\rangle=0, \\
\end{array}
$$
where $|\alpha^{*}\rangle$ is the complex conjugate of
$|\alpha\rangle$. This requires that
$$
\langle\gamma|\mathcal{P}|\gamma\rangle=0\quad \mathrm{and} \quad
\langle\alpha^{*}|\langle\beta|\mathcal{Q}|\alpha^{*}\rangle|\beta\rangle=0.
$$
So the positive operators $\mathcal{P}$ and $\mathcal{Q}$ have to
be orthogonal to all $|\gamma\rangle$'s and
$|\alpha^{*}\rangle\otimes|\beta\rangle$'s resppectively. As it is
shown in appendix II, such $\mathcal{P}$ and $\mathcal{Q}$
actually exist and hence $W'_{1}$ is a d-EW. Similarly, we can
show that EWs (\ref{bwitnessapprox}) are all d-EWs.
\par
In the same way, it can be shown that the remaining boundaries are
also not hyper-planes by itself and the EWs corresponding to
hyper-planes approximating them, i.e.,
\begin{equation}\label{b2witnessapprox}
\begin{array}{c}
  W_{2}=2I_{_{3}}\otimes I_{_{3}}-\frac{3}{2}[(-1)^{i_{_{1}}}
    \lambda_{_{1}}\otimes\lambda_{_{1}}+(-1)^{i_{_{2}}}\lambda_{_{2}}\otimes\lambda_{_{2}}
    -\lambda_{_{3}}\otimes\lambda_{_{3}}+(-1)^{i_{_{4}}}\lambda_{_{4}}\otimes\lambda_{_{4}} \\
  +(-1)^{i_{_{5}}}\lambda_{_{5}}\otimes\lambda_{_{5}}
    +(-1)^{i_{_{6}}}\lambda_{_{6}}\otimes\lambda_{_{6}}
    +(-1)^{i_{_{7}}}\lambda_{_{7}}\otimes\lambda_{_{7}}-\lambda_{_{8}}\otimes\lambda_{_{8}}],\\
\end{array}
\end{equation}
and
\begin{equation}\label{b3witnessapprox}
\begin{array}{c}
  W_{3}=\frac{11}{8}I_{_{3}}\otimes I_{_{3}}-\frac{3}{4}[(-1)^{i_{_{1}}}
    \lambda_{_{1}}\otimes\lambda_{_{1}}+(-1)^{i_{_{2}}}\lambda_{_{2}}\otimes\lambda_{_{2}}
    -\lambda_{_{3}}\otimes\lambda_{_{3}}+\lambda_{_{8}}\otimes\lambda_{_{8}}] \\
  \hspace{1cm}-\frac{9}{8}[(-1)^{i_{_{4}}}\lambda_{_{4}}\otimes\lambda_{_{4}}+(-1)^{i_{_{5}}}\lambda_{_{5}}\otimes\lambda_{_{5}}
    +(-1)^{i_{_{6}}}\lambda_{_{6}}\otimes\lambda_{_{6}}+(-1)^{i_{_{7}}}\lambda_{_{7}}\otimes\lambda_{_{7}}], \\
\end{array}
\end{equation}
are all d-EWs.
\par
In summary, the EWs of the form (\ref{ew}) which correspond to
boundaries or approximated boundaries hyper-planes of the feasible
region, are all d-EWs. The other EWs of the form (\ref{ew}) can be
written as a convex combination of a positive operator and one of
the EWs coming from boundary hyper-planes \cite{jafar4}. So if all
the boundaries would be hyper-planes by itself, we could say that
$\lambda$-diagonal EWs are all decomposable. Therefore, it is
tempting to add some off-diagonal terms
$\lambda_{_{i}}\otimes\lambda_{_{j}}$ with $i\neq j$ to the
operator of (\ref{ew}) in order to get nd-EWs which is the topic
of the next section.
\section{$\lambda$-non-diagonal EWs}
\subsection{First category}
For the first ctegory of $\lambda$-non-diagonal EWs, let us
consider a Hermitian operator of the form
\begin{equation}\label{ew1}
\begin{array}{c}
  \mathcal{W}=a_{_{0}}I_{_{3}}\otimes I_{_{3}}+\sum_{k=1}^8
a_{_{k}}\lambda_{_{k}}\otimes\lambda_{_{k}}+
a_{_{1,2}}\lambda_{_{1}}\otimes\lambda_{_{2}}+a_{_{2,1}}\lambda_{_{2}}\otimes\lambda_{_{1}} \\
  \hspace{1cm}+a_{_{4,5}}\lambda_{_{4}}\otimes\lambda_{_{5}}+a_{_{5,4}}\lambda_{_{5}}\otimes\lambda_{_{4}}
+a_{_{6,7}}\lambda_{_{6}}\otimes\lambda_{_{7}}+a_{_{7,6}}\lambda_{_{7}}\otimes\lambda_{_{6}} \\
\end{array}
\end{equation}
Vertices of the feasible region coming from pure product states
are listed in the table 2.
\begin{table}[h]
\renewcommand{\arraystretch}{1}
\addtolength{\arraycolsep}{-2pt}
$$
\small{
\begin{array}{cccccccccccccc|c}\hline
  P_{1} & P_{2} & P_{3} & P_{4} & P_{5} & P_{6} & P_{7} & P_{8} & P_{1,2} & P_{2,1} & P_{4,5} & P_{5,4} & P_{6,7} & P_{7,6} & \mathrm{Product \ state}
  \\ \hline
  \pm1 & 0 & 0 & 0 & 0 & 0 & 0 & \frac{1}{3} & 0 & 0 & 0 & 0 & 0 & 0 &
  |\lambda_{_{1}};+1\rangle|\lambda_{_{1}};\pm1\rangle\\
  0 & \pm1 & 0 & 0 & 0 & 0 & 0 & \frac{1}{3} & 0 & 0 & 0 & 0 & 0 & 0 & |\lambda_{_{2}};+1\rangle|\lambda_{_{2}};\pm1\rangle \\
  0 & 0 & \pm1 & 0 & 0 & 0 & 0 & \frac{1}{3} & 0 & 0 & 0 & 0 & 0 & 0 & |\lambda_{_{3}};+1\rangle|\lambda_{_{3}};\pm1\rangle \\
  0 & 0 & 0 & 0 & 0 & 0 & 0 & \frac{1}{3} & \pm1 & 0 & 0 & 0 & 0 & 0 & |\lambda_{_{1}};+1\rangle|\lambda_{_{2}};\pm1\rangle \\
  0 & 0 & 0 & 0 & 0 & 0 & 0 & \frac{1}{3} & 0 & \pm1 & 0 & 0 & 0 & 0 & |\lambda_{_{2}};+1\rangle|\lambda_{_{1}};\pm1\rangle \\
  0 & 0 & \frac{1}{4} & \pm1 & 0 & 0 & 0 & \frac{1}{12} & 0 & 0 & 0 & 0 & 0 & 0 & |\lambda_{_{4}};+1\rangle|\lambda_{_{4}};\pm1\rangle \\
  0 & 0 & \frac{1}{4} & 0 & \pm1 & 0 & 0 & \frac{1}{12}& 0 & 0 & 0 & 0 & 0 & 0 & |\lambda_{_{5}};+1\rangle|\lambda_{_{5}};\pm1\rangle \\
  0 & 0 & \frac{1}{4} & 0 & 0 & \pm1 & 0 & \frac{1}{12} & 0 & 0 & 0 & 0 & 0 & 0 & |\lambda_{_{6}};+1\rangle|\lambda_{_{6}};\pm \rangle \\
  0 & 0 & \frac{1}{4} & 0 & 0 & 0 & \pm1 & \frac{1}{12} & 0 & 0 & 0 & 0 & 0 & 0 & |\lambda_{_{7}};+1\rangle|\lambda_{_{7}};\pm \rangle \\
  0 & 0 & \frac{1}{4} & 0 & 0 & 0 & 0 & \frac{1}{12} & 0 & 0 & \pm1 & 0 & 0 & 0 & |\lambda_{_{4}};+1 \rangle|\lambda_{_{5}};\pm \rangle \\
  0 & 0 & \frac{1}{4} & 0 & 0 & 0 & 0 & \frac{1}{12} & 0 & 0 & 0 & \pm1 & 0 & 0 & |\lambda_{_{5}};+1\rangle|\lambda_{_{4}};\pm \rangle\\
  0 & 0 & \frac{1}{4} & 0 & 0 & 0 & 0 & \frac{1}{12} & 0 & 0 & 0 & 0 & \pm1 & 0 & |\lambda_{_{6}};+1 \rangle|\lambda_{_{7}};\pm \rangle\\
  0 & 0 & \frac{1}{4} & 0 & 0 & 0 & 0 & \frac{1}{12} & 0 & 0 & 0 & 0 & 0 & \pm1 & |\lambda_{_{7}};+1\rangle|\lambda_{_{6}};\pm \rangle \\
  0 & 0 & 0 & 0 & 0 & 0 & 0 & \frac{4}{3} & 0 & 0 & 0 & 0 & 0 & 0 & |\lambda_{_{8}};\frac{-2}{\sqrt{3}}\rangle|\lambda_{_{8}};\frac{-2}{\sqrt{3}}\rangle \\
  0 & 0 & 0 & 0 & 0 & 0 & 0 & -\frac{2}{3} & 0 & 0 & 0 & 0 & 0 & 0 & |\lambda_{_{8}};\frac{1}{\sqrt{3}}\rangle|\lambda_{_{8}};\frac{-2}{\sqrt{3}}\rangle
  \\ \hline
\end{array}}
$$
\caption{The product vectors and coordinates of vertices for
$\mathcal{W}$.}
\renewcommand{\arraystretch}{1}
\addtolength{\arraycolsep}{-3pt}
\end{table}

The hyper-plane passing through the two vertices
$$
(0,0,0,0,0,0,0,\frac{4}{3},,0,0,0,0,0,0)\quad,\quad
(0,0,1,0,0,0,0,\frac{1}{3},0,0,0,0,0,0)
$$
along with all other
vertices having a coordinate +1, is
\begin{equation}\label{boffplane1}
           \frac{3}{4}(P_{_{1}}+P_{_{2}}+P_{_{1,2}}+P_{_{2,1}}+P_{_{3}}+P_{_{4}}+P_{_{5}}
       +P_{_{4,5}}+P_{_{5,4}}+P_{_{6}}+P_{_{7}}+P_{_{6,7}}+P_{_{7,6}}+P_{_{8}})-1=0.
\end{equation}
As the proof (see appendix II) shows, there exist pure product
states that maximizes the left hand side of (\ref{boffplane1}) and
the maximum value is $\frac{3}{4}$, greater than zero. So this
hyper-plane can not be a real boundary of the feasible region and
we need to approximate the corresponding real boundary with the
following hyper-plane parallel to it
\begin{equation}\label{apboffplane1}
    \frac{3}{4}(P_{_{1}}+P_{_{2}}+P_{_{1,2}}+P_{_{2,1}}+P_{_{3}}+P_{_{4}}+P_{_{5}}
       +P_{_{4,5}}+P_{_{5,4}}+P_{_{6}}+P_{_{7}}+P_{_{6,7}}+P_{_{7,6}}+P_{_{8}})-\frac{7}{4}=0.
\end{equation}
The corresponding EW of hyper-plane (\ref{apboffplane1}) is as
follows
\begin{equation}\label{apoffwit1}
    \begin{array}{c}
  \mathcal{W}_{1}=\frac{7}{4}I_{_{3}}\otimes I_{_{3}}-\frac{3}{4}(\sum_{k=1}^8
\lambda_{_{k}}\otimes\lambda_{_{k}}+
\lambda_{_{1}}\otimes\lambda_{_{2}}+\lambda_{_{2}}\otimes\lambda_{_{1}} \\
  \hspace{1cm}+\lambda_{_{4}}\otimes\lambda_{_{5}}+\lambda_{_{5}}\otimes\lambda_{_{4}}
+\lambda_{_{6}}\otimes\lambda_{_{7}}+\lambda_{_{7}}\otimes\lambda_{_{6}}) \\
\end{array}
\end{equation}
$\mathcal{W}_{1}$ is a nd-EW since it is able to detect a bound
entangled state. To show this, we note that the density matrix
\begin{equation}\label{}
    \begin{array}{c}
       \rho=\frac{1}{9}I_{_{3}}\otimes
I_{_{3}}+\frac{c}{6(a+2c)}(\lambda_{_{1}}\otimes\lambda_{_{1}}
+\lambda_{_{2}}\otimes\lambda_{_{2}}+\lambda_{_{1}}\otimes\lambda_{_{2}}
+\lambda_{_{2}}\otimes\lambda_{_{1}}+\lambda_{_{4}}\otimes\lambda_{_{4}}
+\lambda_{_{5}}\otimes\lambda_{_{5}}+\lambda_{_{4}}\otimes\lambda_{_{5}}\\
       +\lambda_{_{5}}\otimes\lambda_{_{4}}
+\lambda_{_{6}}\otimes\lambda_{_{6}}+\lambda_{_{7}}\otimes\lambda_{_{7}}
+\lambda_{_{6}}\otimes\lambda_{_{7}}+\lambda_{_{7}}\otimes\lambda_{_{6}})
+\frac{a-c}{6(a+2c)}(\lambda_{_{3}}\otimes\lambda_{_{3}}+\lambda_{_{8}}\otimes\lambda_{_{8}})\\
     \end{array}
\end{equation}
where $a$ and $c$ are non-negative real parameters such that
$0\leq c\leq \frac{a}{\sqrt{3}}$ is a PPT state. On the other
hand,
$$
Tr(\mathcal{W}_{1}\rho)=\frac{3}{4}\ \frac{a-2c}{a+2c}.
$$
It is seen that for $c>\frac{a}{2}$ the PPT state $\rho$ can be
detected by $\mathcal{W}_{1}$. Hence, for $\frac{a}{2}<c\leq
\frac{a}{\sqrt{3}}$, $\rho$ is a bound entangled state and
$\mathcal{W}_{1}$ is a nd-EW.
\par
This category contains $2^{14}$ EWs corresponding to $2^{14}$
exact or approximated boundary hyper-planes of the feasible
region. This set of EWs can be partitioned into $64$ equivalence
classes of size $2^{8}$ by the conjugation action of the exchange
operator
\begin{equation}\label{exop}
 \Pi=\frac{1}{3}I_{_{3}}\otimes I_{_{3}}+\frac{1}{2}\sum_{k=1}^8
\lambda_{_{k}}\otimes\lambda_{_{k}}
\end{equation}
which exchanges the two parties, and the operators $T\otimes
I_{_{3}}$, and $M_{_{i}}\otimes I_{_{3}},\; i=1,2,3$ on it, where
T is the transpose operator and
\begin{equation}\label{symop1}
    M_{_{1}}=\left(%
\begin{array}{ccc}
  i & 0 & 0 \\
  0 & 1 & 0 \\
  0 & 0 & -1 \\
\end{array}%
\right)\quad,\quad M_{_{2}}=\left(%
\begin{array}{ccc}
  1 & 0 & 0 \\
  0 & i & 0 \\
  0 & 0 & -1 \\
\end{array}%
\right)\quad,\quad M_{_{3}}=\left(%
\begin{array}{ccc}
  1 & 0 & 0 \\
  0 & -1 & 0 \\
  0 & 0 & i \\
\end{array}%
\right).
\end{equation}
Any two EWs are said to be equivalent if they can be transformed
into each other by the above five operators or their possible
products. The operators $M_{_{i}},\; i=1,2,3$ do not change
Gell-Mann matrices $\lambda_{_{3}}$ and $\lambda_{_{8}}$ while
induce the following transformations on the other Gell-Mann
matrices
\begin{equation}\label{}
\begin{array}{c}
M_{_{1}} :\hspace{.2cm}\lambda_{_{1}}\rightarrow-\lambda_{_{2}},
\quad\lambda_{_{2}}\rightarrow\lambda_{_{1}},
\quad\lambda_{_{4}}\rightarrow\lambda_{_{5}},\quad
\lambda_{_{5}}\rightarrow-\lambda_{_{4}},
\quad\lambda_{_{6}}\rightarrow-\lambda_{_{6}},
\quad\lambda_{_{7}}\rightarrow-\lambda_{_{7}}, \\
M_{_{2}} :\hspace{.2cm}\lambda_{_{1}}\rightarrow\lambda_{_{2}},
\quad\lambda_{_{2}}\rightarrow-\lambda_{_{1}},
\quad\lambda_{_{4}}\rightarrow-\lambda_{_{4}},\quad
\lambda_{_{5}}\rightarrow-\lambda_{_{5}},
\quad\lambda_{_{6}}\rightarrow\lambda_{_{7}},
\quad\lambda_{_{7}}\rightarrow-\lambda_{_{6}}, \\
M_{_{3}} :\hspace{.2cm}\lambda_{_{1}}\rightarrow-\lambda_{_{1}},
\quad\lambda_{_{2}}\rightarrow-\lambda_{_{2}},
\quad\lambda_{_{4}}\rightarrow\lambda_{_{5}},\quad
\lambda_{_{5}}\rightarrow-\lambda_{_{4}},
\quad\lambda_{_{6}}\rightarrow-\lambda_{_{7}},
\quad\lambda_{_{7}}\rightarrow\lambda_{_{6}}. \\
\end{array}
\end{equation}
We have discussed one of the equivalence classes. As the
discussion of the other classes is similar, we do not discuss them
here.
\subsection{Second category}
For the second category of $\lambda$-non-diagonal EWs, let us
consider the following Hermitian operator
\begin{equation}\label{ew2}
\begin{array}{c}
  \mathcal{W'}=a_{_{0}}I_{_{3}}\otimes I_{_{3}}+\sum_{k=1}^8
a_{_{k}}\lambda_{_{k}}\otimes\lambda_{_{k}}+
a_{_{3,8}}\sqrt{3}\lambda_{_{3}}\otimes\lambda_{_{8}}+a_{_{8,3}}\sqrt{3}\lambda_{_{8}}\otimes\lambda_{_{3}}.
\end{array}
\end{equation}
Vertices of the feasible region coming from pure product states
are listed in the table 3 where $
P_{3,8}=Tr(\sqrt{3}\lambda_{_{3}}\otimes\lambda_{_{8}}|\gamma\rangle\langle\gamma|)
$ and $P_{8,3}$ is defined similarly.
\begin{table}[h]
\renewcommand{\arraystretch}{1}
\addtolength{\arraycolsep}{-2pt}
$$
\small{
\begin{array}{cccccccccc|c}\hline
  P_{1} & P_{2} & P_{3} & P_{4} & P_{5} & P_{6} & P_{7} & P_{8} & P_{3,8} & P_{8,3}& \mathrm{Product \ state}
  \\ \hline
  \pm1 & 0 & 0 & 0 & 0 & 0 & 0 & \frac{1}{3} & 0 & 0 &
  |\lambda_{_{1}};+1\rangle|\lambda_{_{1}};\pm1\rangle\\
  0 & \pm1 & 0 & 0 & 0 & 0 & 0 & \frac{1}{3} & 0 & 0 & |\lambda_{_{2}};+1\rangle|\lambda_{_{2}};\pm1\rangle\\
  0 & 0 & +1 & 0 & 0 & 0 & 0 & \frac{1}{3} & \pm1 & \pm1 & |\lambda_{_{3}};\pm1\rangle|\lambda_{_{3}};\pm1\rangle \\
  0 & 0 & -1 & 0 & 0 & 0 & 0 & \frac{1}{3} & \pm1 & \mp1 & |\lambda_{_{3}};\pm1\rangle|\lambda_{_{3}};\mp1\rangle \\
  0 & 0 & \frac{1}{4} & \pm1 & 0 & 0 & 0 & \frac{1}{12} & -\frac{1}{4} & -\frac{1}{4} & |\lambda_{_{4}};+1\rangle|\lambda_{_{4}};\pm1\rangle \\
  0 & 0 & \frac{1}{4} & 0 & \pm1 & 0 & 0 & \frac{1}{12}& -\frac{1}{4} & -\frac{1}{4} & |\lambda_{_{5}};+1\rangle|\lambda_{_{5}};\pm1\rangle \\
  0 & 0 & \frac{1}{4} & 0 & 0 & \pm1 & 0 & \frac{1}{12} & \frac{1}{4} & \frac{1}{4} & |\lambda_{_{6}};+1\rangle|\lambda_{_{6}};\pm \rangle \\
  0 & 0 & \frac{1}{4} & 0 & 0 & 0 & \pm1 & \frac{1}{12} & \frac{1}{4} & \frac{1}{4} & |\lambda_{_{7}};+1\rangle|\lambda_{_{7}};\pm \rangle \\
  0 & 0 & 0 & 0 & 0 & 0 & 0 & \frac{4}{3} & 0 & 0 & |\lambda_{_{8}};\frac{-2}{\sqrt{3}}\rangle|\lambda_{_{8}};\frac{-2}{\sqrt{3}}\rangle \\
  0 & 0 & 0 & 0 & 0 & 0 & 0 & -\frac{2}{3} & \pm2 & 0 &|\lambda_{_{8}};\frac{1}{\sqrt{3}}\rangle|\lambda_{_{8}};\frac{-2}{\sqrt{3}}\rangle\\
  0 & 0 & 0 & 0 & 0 & 0 & 0 & -\frac{2}{3} & 0 & \pm2 &|\lambda_{_{8}};\frac{-2}{\sqrt{3}}\rangle|\lambda_{_{3}};\frac{1}{\sqrt{3}}\rangle
  \\ \hline
\end{array}}
$$
\caption{The product vectors and coordinates of vertices for
$\mathcal{W'}$.}
\renewcommand{\arraystretch}{1}
\addtolength{\arraycolsep}{-3pt}
\end{table}
To show the efficiency of the present method, we try to construct
nd-EWs which are capable to detect the well-known one parameter
family of bound entangled states
\begin{equation}\label{}
\begin{array}{c}
  \rho_{_{b}}=\frac{1}{9}I_{_{3}}\otimes I_{_{3}}+\frac{1}{21}(\lambda_{_{1}}\otimes\lambda_{_{1}}-\lambda_{_{2}}\otimes\lambda_{_{2}}
    +\lambda_{_{4}}\otimes\lambda_{_{4}}-\lambda_{_{5}}\otimes\lambda_{_{5}}
    +\lambda_{_{6}}\otimes\lambda_{_{6}}-\lambda_{_{7}}\otimes\lambda_{_{7}})\\
    -\frac{1}{84}(\lambda_{_{3}}\otimes\lambda_{_{3}}+\lambda_{_{8}}\otimes\lambda_{_{8}})
    -\frac{\sqrt{3}}{84}(5-2b)(\lambda_{_{3}}\otimes\lambda_{_{8}}-\lambda_{_{8}}\otimes\lambda_{_{3}}),\quad\quad 0\leq b\leq5,\\
 \end{array}
\end{equation}
introduced in \cite{horod1}. The Horodecki states $\rho_{_{b}}$
are PPT for $1 \leq b \leq 4$ and as it was shown in
\cite{horod1,doherty3}, the states are separable for $2 \leq b
\leq 3$ and bound entangled for  $1\leq b<2$ and $3 < b \leq 4$.
In \cite{doherty3}, Doherty and et al. introduced an nd-EW which
detects the Horodecki bound entangled states $\rho_{_{b}}$ for
$3<b\leq4$.
\par
Let us proceed to construct an nd-EW which detects $\rho_{_{b}}$
for $3<b\leq4$. To this aim, we begin with the following ten
vertices of the feasible region \small{$$ \begin{array}{ccc}
  (1,0,0,0,0,0,0,\frac{1}{3},0,0), & (0,-1,0,0,0,0,0,\frac{1}{3},0,0), & (0,0,-1,0,0,0,0,\frac{1}{3},1,-1), \\
  (0,0,\frac{1}{4},1,0,0,0,\frac{1}{12},-\frac{1}{4},-\frac{1}{4}), & (0,0,\frac{1}{4},0,-1,0,0,\frac{1}{12},
  -\frac{1}{4},-\frac{1}{4}), & (0,0,\frac{1}{4},0,0,1,0,\frac{1}{12},\frac{1}{4},\frac{1}{4}), \\
 (0,0,\frac{1}{4},0,0,0,-1,\frac{1}{12},\frac{1}{4},\frac{1}{4}), & (0,0,0,0,0,0,0,-\frac{2}{3},2,0),
  &(0,0,0,0,0,0,0,-\frac{2}{3},0,-2), \\
\ & (0,0,0,0,0,0,0,\frac{4}{3},0,0). & \ \\
\end{array}
$$}
The hyper-plane passing through these vertices has the form
$$
\frac{3}{4}(P_{_{1}}-P_{_{2}}+P_{_{4}}-P_{_{5}}+P_{_{6}}-P_{_{7}})
+\frac{3}{4}(P_{_{3}}+P_{_{8}})+\frac{3\sqrt{3}}{4}(P_{_{3,8}}-P_{_{8,3}})-1=0.
$$
Some calculations show that the maximum value of the left hand
side is $\frac{3}{8}$. The pure product state giving this maximum
value lies on the boundary of the feasible region and we replace
its corresponding point with one of the above ten vertices. Then
we pass a hyper-plane through the second set of ten points and
find pure product states maximizing it. Again, we replace the
corresponding points of these states with suitable points of the
second set. By continuing this process, we end up with ten
vertices of the table 4.
\begin{table}[h]
\renewcommand{\arraystretch}{1}
\addtolength{\arraycolsep}{-2pt}
$$
\small{
\begin{array}{cccccccccc|c}\hline
  P_{1} & P_{2} & P_{3} & P_{4} & P_{5} & P_{6} & P_{7} & P_{8} & P_{3,8} & P_{8,3}& \mathrm{Product \ state}
  \\ \hline
  0 & 0 & -1 & 0 & 0 & 0 & 0 & \frac{1}{3} & 1 & -1 &
 \tiny{\left(%
\begin{array}{c}
  1 \\
  0 \\
  0 \\
\end{array}
\right)}\otimes\tiny{\left(%
\begin{array}{c}
  0 \\
  1 \\
  0 \\
\end{array}
\right)}\\
  0 & 0 & 0 & 0 & 0 & 0 & 0 &-\frac{2}{3} & 2 & 0 & \tiny{\left(%
\begin{array}{c}
  0 \\
  1 \\
  0 \\
\end{array}
\right)}\otimes\tiny{\left(%
\begin{array}{c}
  0 \\
  0 \\
  1 \\
\end{array}
\right)}\\
  0 & 0 & 0 & 0 & 0 & 0 & 0 & -\frac{2}{3} & 0 & -2 & \tiny{\left(%
\begin{array}{c}
  0 \\
  0 \\
  1 \\
\end{array}
\right)}\otimes\tiny{\left(%
\begin{array}{c}
  1 \\
  0 \\
  0 \\
\end{array}
\right)}\\
  \frac{4}{9} & 0 & 0 & \frac{4}{9} & 0 & \frac{4}{9} & 0 & 0 & 0 & 0 &
  \tiny{\frac{1}{\sqrt{3}}\left(%
\begin{array}{c}
  1 \\
  1 \\
  1 \\
\end{array}
\right)}\otimes\tiny{\frac{1}{\sqrt{3}}\left(%
\begin{array}{c}
  1 \\
  1 \\
  1 \\
\end{array}
\right)}\\
  \frac{4}{9} & 0 & 0 & \frac{1}{9} & -\frac{1}{3} & \frac{1}{9} & -\frac{1}{3} & 0 & 0 & 0 &
  \tiny{\frac{1}{\sqrt{3}}\left(%
\begin{array}{c}
  1 \\
  1 \\
  \omega \\
\end{array}
\right)}\otimes\tiny{\frac{1}{\sqrt{3}}\left(%
\begin{array}{c}
  1 \\
  1 \\
  \bar{\omega} \\
\end{array}
\right)}\\
  0 & -\frac{4}{9} & 0 & 0 & -\frac{4}{9} & \frac{4}{9} & 0 & 0 & 0 & 0 &
  \tiny{\frac{1}{\sqrt{3}}\left(%
\begin{array}{c}
  i \\
  1 \\
  1\\
\end{array}
\right)}\otimes\tiny{\frac{1}{\sqrt{3}}\left(%
\begin{array}{c}
  -i \\
  1 \\
  1 \\
\end{array}
\right)}\\
  0 & -\frac{4}{9} & 0 & \frac{4}{9} & 0 & 0 & -\frac{4}{9} & 0 & 0 & 0 &
  \tiny{\frac{1}{\sqrt{3}}\left(%
\begin{array}{c}
  1 \\
  i \\
  1\\
\end{array}
\right)}\otimes\tiny{\frac{1}{\sqrt{3}}\left(%
\begin{array}{c}
  1 \\
  -i \\
  1 \\
\end{array}
\right)}\\
  0 & 0 & \frac{3}{16} & 0 & 0 & \frac{3}{4} & 0 & -\frac{5}{48} & \frac{15}{16} & -\frac{1}{16} &
  \tiny{\frac{1}{2}\left(%
\begin{array}{c}
  0 \\
  \sqrt{3} \\
  1\\
\end{array}
\right)}\otimes\tiny{\frac{1}{2}\left(%
\begin{array}{c}
  0 \\
  1 \\
  \sqrt{3} \\
\end{array}
\right)}\\
  \frac{64}{225} & 0 & 0 & \frac{112}{225} & 0 & \frac{112}{225} & 0 & \frac{4}{75} & 0 & 0 &
  \tiny{\frac{1}{\sqrt{15}}\left(%
\begin{array}{c}
  2 \\
  2 \\
  \sqrt{7}\\
\end{array}
\right)}\otimes\tiny{\frac{1}{\sqrt{15}}\left(%
\begin{array}{c}
  2 \\
  2 \\
  \sqrt{7} \\
\end{array}
\right)}\\
  \frac{2464}{9025} & 0 & \frac{81}{1444} & \frac{2592}{9025} & 0 & \frac{6237}{9025} & 0 &
  \frac{2809}{108300} & \frac{477}{722
  0} & \frac{477}{7220} &
  \tiny{\frac{1}{\sqrt{190}}\left(%
\begin{array}{c}
  \sqrt{32} \\
  \sqrt{77} \\
  9\\
\end{array}
\right)}\otimes\tiny{\frac{1}{\sqrt{190}}\left(%
\begin{array}{c}
  \sqrt{32} \\
  \sqrt{77} \\
  9 \\
\end{array}
\right)}\\
     \hline
\end{array}}
$$
\caption{\small{The product vectors and coordinates of vertices
for $\mathcal{W'}_{_{1}}$. Here $\omega=\exp(\frac{2i\pi}{3})$ and
$i=\sqrt{-1}$.}}
\renewcommand{\arraystretch}{1}
\addtolength{\arraycolsep}{-3pt}
\end{table}
An analogous argument as above give the EW
\begin{equation}\label{dew1}
\begin{array}{c}
  \mathcal{W'}_{_{1}}=\frac{809}{790}I_{_{3}}\otimes I_{_{3}}
  -\frac{1}{6320}[2553(\lambda_{_{1}}\otimes\lambda_{_{1}}-\lambda_{_{2}}\otimes\lambda_{_{2}})
    +5227(\lambda_{_{4}}\otimes\lambda_{_{4}}-\lambda_{_{5}}\otimes\lambda_{_{5}})]
    -\frac{161}{158}(\lambda_{_{6}}\otimes\lambda_{_{6}}-\lambda_{_{7}}\otimes\lambda_{_{7}})\\
    +\frac{501}{790}(\lambda_{_{3}}\otimes\lambda_{_{3}}+\lambda_{_{8}}\otimes\lambda_{_{8}})
    -\frac{114\sqrt{3}}{395}(\lambda_{_{3}}\otimes\lambda_{_{8}}-\lambda_{_{8}}\otimes\lambda_{_{3}}).\\
 \end{array}
\end{equation}
which corresponds to the hyper-plane tangent to the feasible
region and parallel to the hyper-plane passing through the
vertices of the table 4. The nd-EW $\mathcal{W'}_{_{1}}$ detects
the Horodecki states $\rho_{_{b}}$ for
$\frac{2869}{912}(\simeq3.146)<b\leq4$. This nd-EW is very similar
to the nd-EW introduced in \cite{doherty3}.
\par
To construct another nd-EW which detects $\rho_{_{b}}$ for $1\leq
b<2$, we replace the first three points and the 8th point of the
table 4 with the three points
$$
(0,0,-1,0,0,0,0,\frac{1}{3},-1,1),\quad(0,0,0,0,0,0,0,-\frac{2}{3},-2,0),\quad(0,0,0,0,0,0,0,-\frac{2}{3},0,2),
$$
of the table 3 and the point
$(0,0,\frac{3}{16},0,0,\frac{3}{4},0,-\frac{5}{48},-\frac{15}{16},\frac{1}{16})$
respectively. The last point corresponds to a pure product state
coming from interchanging the states of two parties in the pure
product state of the 8th point of the table 4. The hyper-plane
tangent to the feasible region and parallel to the hyper-plane
passing through the new ten points, corresponds to the EW
\begin{equation}\label{dew2}
\begin{array}{c}
  \mathcal{W'}_{_{2}}=\frac{809}{790}I_{_{3}}\otimes I_{_{3}}
  -\frac{1}{6320}[2553(\lambda_{_{1}}\otimes\lambda_{_{1}}-\lambda_{_{2}}\otimes\lambda_{_{2}})
    +5227(\lambda_{_{4}}\otimes\lambda_{_{4}}-\lambda_{_{5}}\otimes\lambda_{_{5}})]
    -\frac{161}{158}(\lambda_{_{6}}\otimes\lambda_{_{6}}-\lambda_{_{7}}\otimes\lambda_{_{7}})\\
    +\frac{501}{790}(\lambda_{_{3}}\otimes\lambda_{_{3}}+\lambda_{_{8}}\otimes\lambda_{_{8}})
    +\frac{114\sqrt{3}}{395}(\lambda_{_{3}}\otimes\lambda_{_{8}}-\lambda_{_{8}}\otimes\lambda_{_{3}}).\\
 \end{array}
\end{equation}
The nd-EW $\mathcal{W'}_{_{2}}$ detects the Horodecki states
$\rho_{_{b}}$ for $1\leq b<\frac{89}{48}(\simeq 1.854)$.
\par
It is noted that the EW (\ref{dew2}) can be obtained from EW
(\ref{dew1}) by the exchange operator $\Pi$ defined in
(\ref{exop}), i.e.,
$$
\mathcal{W'}_{_{2}}=\Pi\;\mathcal{W'}_{_{1}}\Pi^{\dagger}.
$$
In fact, this category contains $13\times2^{6}$ EWs corresponding
to $5\times2^{6}$ exact or approximated boundary hyper-planes of
the feasible region. This set of EWs can be partitioned into $36$
equivalence classes, 20 classes of size 8 and 16 classes of size
16, by the conjugation action of exchange operator $\Pi$ and the
operators $T\otimes I_{_{3}}$, $M_{_{1}}^{^{2}}\otimes I_{_{3}}$
and $M_{_{2}}^{^{2}}\otimes I_{_{3}}$ on it, where T is the
transpose operator and $M_{_{1}}$ and $M_{_{2}}$ are operators
defined in (\ref{symop1}). Any two EWs are said to be equivalent
if they can be transformed into each other by the above four
operators or their possible products. So we have studied one of
the equivalence classes of size 16 which contains the two EWs
$\mathcal{W'}_{_{1}}$ and $\mathcal{W'}_{_{2}}$. The other classes
can be studied similarly.
\section{Conclusion}
We have considered the EWs that can be constructed from Gell-Mann
matrices for Lie algebra su(3) by using the Linear programming
method. The general form of $\lambda$-diagonal and two cases of
$\lambda$-non-diagonal EWs have been discussed in detail. It has
been shown that in all considered cases the feasible region is a
polygon approximately. In $\lambda$-diagonal case, the boundaries
of the polygon correspond to d-EWs or positive operators which in
turn implies that up to our approximation, all $\lambda$-diagonal
EWs are decomposable. However, in $\lambda$-non-diagonal case, a
large number of boundaries of the polygon correspond to nd-EWs.
The decomposability of presented $\lambda$-diagonal EWs has been
proved by writing them explicitly in decomposable form or by using
the pure product states having zero expectation values with them.
To show the non-decomposability of some EWs in
$\lambda$-non-diagonal case, we have constructed a class of bound
entangled states or used known bound entangled states that can be
detected by them. Although the method may be tedious but as it was
shown in the case of Horodecki bound entangled states, it clarify
that in principle one can construct suitable
$\lambda$-non-diagonal EWs for any two-qutrit bound entangled
state.
\par
Since the boundaries of the feasible region in $\lambda$-diagonal
EWs are not all hyper-planes by itself, it remains as an open
problem to show that $\lambda$-diagonal EWs are all decomposable.

\vspace{1cm} \setcounter{section}{0}
 \setcounter{equation}{0}
 \renewcommand{\theequation}{I-\arabic{equation}}
\newpage
{\Large{Appendix I:}}\\
\textbf{Lie algebra su(3)}\\
The real Lie algebra su(3) is the algebra of the anti-Hermitian
$3\times 3$ matrices with vanishing trace. The standard Hermitian
basis of su(3) consists of the Gell-Mann $\lambda$ matrices:
$$
\begin{array}{ccc}
 \lambda_{_{1}}= \left(%
\begin{array}{ccc}
  0 & 1 & 0 \\
  1 & 0 & 0 \\
  0 & 0 & 0 \\
\end{array}%
\right) &\lambda_{_{2}}= \left(%
\begin{array}{ccc}
  0 & -i & 0 \\
  i & 0 & 0 \\
  0 & 0 & 0 \\
\end{array}%
\right) & \lambda_{_{3}}= \left(%
\begin{array}{ccc}
  1 & 0 & 0\\
  0 & -1 & 0 \\
  0 & 0 & 0 \\
\end{array}%
\right) \\
 \lambda_{_{4}}= \left(%
\begin{array}{ccc}
  0 & 0 & 1 \\
  0 & 0 & 0 \\
  1 & 0 & 0 \\
\end{array}%
\right) & \lambda_{_{5}}= \left(%
\begin{array}{ccc}
  0 & 0 & -i \\
  0 & 0 & 0 \\
  i & 0 & 0 \\
\end{array}%
\right) & \lambda_{_{6}}= \left(%
\begin{array}{ccc}
  0 & 0 & 0 \\
  0 & 0 & 1 \\
  0 & 1 & 0 \\
\end{array}%
\right) \\
 \lambda_{_{7}}= \left(%
\begin{array}{ccc}
  0 & 0 & 0 \\
  0 & 0 & -i \\
  0 & i & 0 \\
\end{array}%
\right) & \lambda_{_{8}}=\frac{1}{\sqrt{3}}\left(%
\begin{array}{ccc}
  1 & 0 & 0 \\
  0 & 1 & 0 \\
  0 & 0 & -2 \\
\end{array}%
\right)  \\
\end{array}
$$
which fulfill the condition
$$
Tr(\lambda_{_{i}}\lambda_{_{j}})=2\;\delta_{_{ij}}
$$
and the squares of expectation values of $\lambda_{_{i}}$'s over
an arbitrary vector $|\alpha\rangle$ sum to $\frac{4}{3}$; i.e.,
\begin{equation}\label{expeciden}
    \sum_{i=1}^{8}(\langle\alpha|\lambda_{_{i}}|\alpha\rangle)^2=\frac{4}{3}.
\end{equation}
Here, for completeness, we give a proof for the above equality.
Let
$|\alpha\rangle=\alpha_{_{0}}|0\rangle+\alpha_{_{1}}|1\rangle+\alpha_{_{2}}|2\rangle
$ be an arbitrary state. Then we can write
\begin{equation}\label{gelleigen}
\begin{array}{cc}
  \langle\alpha|\lambda_{_{1}}|\alpha\rangle=2\ \mathrm{Re}(\alpha_{_{0}}^{*}\alpha_{_{1}})\quad,\quad
  &
   \langle\alpha|\lambda_{_{2}}|\alpha\rangle=2\ \mathrm{Im}(\alpha_{_{0}}^{*}\alpha_{_{1}})\\
  \langle\alpha|\lambda_{_{3}}|\alpha\rangle=|\alpha_{_{0}}|^{2}-|\alpha_{_{1}}|^{2}\quad,\quad &
  \langle\alpha|\lambda_{_{4}}|\alpha\rangle=2\ \mathrm{Re}(\alpha_{_{0}}^{*}\alpha_{_{2}})\\
  \langle\alpha|\lambda_{_{5}}|\alpha\rangle=2\ \mathrm{Im}(\alpha_{_{0}}^{*}\alpha_{_{2}})\quad,\quad &
   \langle\alpha|\lambda_{_{6}}|\alpha\rangle=2\ \mathrm{Re}(\alpha_{_{1}}^{*}\alpha_{_{2}}) \\
  \langle\alpha|\lambda_{_{7}}|\alpha\rangle=2\ \mathrm{Im}(\alpha_{_{1}}^{*}\alpha_{_{2}}) \quad,\quad&
   \langle\alpha|\lambda_{_{8}}|\alpha\rangle=\frac{1}{\sqrt{3}}-\frac{3}{\sqrt{3}}|\alpha_{_{2}}|^{2}\\
\end{array}
\end{equation}
whence
$$
\sum_{i=1}^{8}(\langle\alpha|\lambda_{_{i}}|\alpha\rangle)^2=4|\alpha_{_{0}}|^{2}|\alpha_{_{1}}|^{2}+
(|\alpha_{_{0}}|^{2}-|\alpha_{_{1}}|^{2})^{2}+4|\alpha_{_{0}}|^{2}|\alpha_{_{2}}|^{2}+
4|\alpha_{_{1}}|^{2}|\alpha_{_{2}}|^{2}+\frac{1}{3}(1-3|\alpha_{_{2}}|^{2})^{2}
$$
$$
=(|\alpha_{_{0}}|^{2}+|\alpha_{_{1}}|^{2})^{2}+4|\alpha_{_{2}}|^{2}(|\alpha_{_{0}}|^{2}+|\alpha_{_{1}}|^{2})+
\frac{1}{3}(1-3|\alpha_{_{2}}|^{2})^{2}
$$
$$
=(|\alpha_{_{0}}|^{2}+|\alpha_{_{1}}|^{2})(|\alpha_{_{0}}|^{2}+
|\alpha_{_{1}}|^{2}+4|\alpha_{_{2}}|^{2})+\frac{1}{3}(1-3|\alpha_{_{2}}|^{2})^{2}
$$
noting the normalization condition
$$
|\alpha_{_{0}}|^{2}+|\alpha_{_{1}}|^{2}+|\alpha_{_{2}}|^{2}=1
$$
the above statement can be written as
$$
(1-|\alpha_{_{2}}|^{2})(1+3|\alpha_{_{2}}|^{2})+\frac{1}{3}(1-3|\alpha_{_{2}}|^{2})^{2}=\frac{4}{3}.
$$

{\Large{Appendix II:}}\\
{\bf Proving the equalities}: \\
In this appendix we give the proof of equalities for some
boundary hyper-planes.\\
{\bf The proof of (\ref{halfspaces})}:\\
Since the proofs are similar, we give a proof only for the
following hyper-plane
$$
\frac{3}{4}\sum_{j=1}^{8}P_{_{j}}-1=0.
$$
Let $|\gamma\rangle=|\alpha\rangle\otimes|\beta\rangle$ be an
arbitrary pure product state. Then, by definition, we have
$$
\frac{3}{4}\sum_{j=1}^{8}P_{_{j}}=
\frac{3}{4}\sum_{j=1}^{8}\langle\alpha|\lambda_{_{j}}|\alpha\rangle\langle\beta|\lambda_{_{j}}|\beta\rangle.
$$
Using the Cauchy-Schwartz inequality and (\ref{expeciden}), we get
the result
$$
\leq\frac{3}{4}\left(\sum_{j=1}^{8}(\langle\alpha|\lambda_{_{j}}|\alpha\rangle)^2\right)^{\frac{1}{2}}
\left(\sum_{j=1}^{8}(\langle\beta|\lambda_{_{j}}|\beta\rangle)^2\right)^{\frac{1}{2}}=
\frac{3}{4}\sqrt{\frac{4}{3}}\sqrt{\frac{4}{3}}=1.
$$
{\bf The proof of (\ref{halfspaces1})}:\\
Since the proofs are similar, we give the proof only for the
following one
$$
\frac{3}{2}\left[P_{_{1}}+P_{_{2}}+P_{_{3}}-P_{_{8}}\right]
+\frac{3}{4}\left[P_{_{4}}+P_{_{5}}+P_{_{6}} +P_{_{7}}\right]-1=0.
$$
By definition of $P_{_{i}}$'s and by noting to (\ref{gelleigen}),
the lhs of this equality can be written as
$$
\mathrm{lhs}=\frac{3}{2}[4\mathrm{Re}(\alpha_{_{0}}^{*}\alpha_{_{1}})\mathrm{Re}(\beta_{_{0}}^{*}\beta_{_{1}})
+4\mathrm{Im}(\alpha_{_{0}}^{*}\alpha_{_{1}})\mathrm{Im}(\beta_{_{0}}^{*}\beta_{_{1}})
+(|\alpha_{_{0}}|^{2}-|\alpha_{_{1}}|^{2})(|\beta_{_{0}}|^{2}-|\beta_{_{1}}|^{2})
$$
$$
-\frac{1}{3}(1-3|\alpha_{_{2}}|^{2})(1-3|\beta_{_{2}}|^{2})]
+\frac{3}{4}[4\mathrm{Re}(\alpha_{_{0}}^{*}\alpha_{_{2}})\mathrm{Re}(\beta_{_{0}}^{*}\beta_{_{2}})
4\mathrm{Im}(\alpha_{_{0}}^{*}\alpha_{_{2}})\mathrm{Im}(\beta_{_{0}}^{*}\beta_{_{2}})
$$
$$
+4\mathrm{Re}(\alpha_{_{1}}^{*}\alpha_{_{2}})\mathrm{Re}(\beta_{_{1}}^{*}\beta_{_{2}})
+4\mathrm{Im}(\alpha_{_{1}}^{*}\alpha_{_{2}})\mathrm{Im}(\beta_{_{1}}^{*}\beta_{_{2}})]-1
$$
$$
=-1+6|\alpha_{_{0}}||\alpha_{_{1}}||\beta_{_{0}}||\beta_{_{1}}|\cos(\theta_{_{01}}-\theta'_{_{01}})
+\frac{3}{2}(|\alpha_{_{0}}|^{2}-|\alpha_{_{1}}|^{2})(|\beta_{_{0}}|^{2}-|\beta_{_{1}}|^{2})
$$
$$
-\frac{1}{2}(1-3|\alpha_{_{2}}|^{2})(1-3|\beta_{_{2}}|^{2})
+3|\alpha_{_{0}}||\alpha_{_{2}}||\beta_{_{0}}||\beta_{_{2}}|\cos(\theta_{_{02}}-\theta'_{_{02}})
+3|\alpha_{_{1}}||\alpha_{_{2}}||\beta_{_{1}}||\beta_{_{2}}|\cos(\theta_{_{12}}-\theta'_{_{12}})
$$
$$
\leq-1+6|\alpha_{_{0}}||\alpha_{_{1}}||\beta_{_{0}}||\beta_{_{1}}|
+\frac{3}{2}(|\alpha_{_{0}}|^{2}-|\alpha_{_{1}}|^{2})(|\beta_{_{0}}|^{2}-|\beta_{_{1}}|^{2})
$$
$$
-\frac{1}{2}(1-3|\alpha_{_{2}}|^{2})(1-3|\beta_{_{2}}|^{2})
+3|\alpha_{_{0}}||\alpha_{_{2}}||\beta_{_{0}}||\beta_{_{2}}|
+3|\alpha_{_{1}}||\alpha_{_{2}}||\beta_{_{1}}||\beta_{_{2}}|
$$
$$
=-1+3|\alpha_{_{2}}||\beta_{_{2}}|(|\alpha_{_{0}}||\beta_{_{0}}|+|\alpha_{_{1}}||\beta_{_{1}}|
+|\alpha_{_{2}}||\beta_{_{2}}|)+\frac{3}{2}(|\alpha_{_{0}}||\beta_{_{0}}|+|\alpha_{_{1}}||\beta_{_{1}}|)^{2}
-\frac{3}{2}(|\alpha_{_{0}}||\beta_{_{1}}|-|\alpha_{_{1}}||\beta_{_{0}}|)^{2}
$$
$$
+\frac{3}{2}(|\alpha_{_{2}}|^{2}+|\beta_{_{2}}|^{2}-5|\alpha_{_{2}}|^{2}|\beta_{_{2}}|^{2}-\frac{1}{3})
$$
By using the Cauchy-Schwartz inequality, we have
$$
\leq-1+3|\alpha_{_{2}}||\beta_{_{2}}|+\frac{3}{2}[(|\alpha_{_{0}}||\beta_{_{0}}|+|\alpha_{_{1}}||\beta_{_{1}}|)^{2}
-(|\alpha_{_{0}}||\beta_{_{1}}|-|\alpha_{_{1}}||\beta_{_{0}}|)^{2}
+|\alpha_{_{2}}|^{2}+|\beta_{_{2}}|^{2}-5|\alpha_{_{2}}|^{2}|\beta_{_{2}}|^{2}-\frac{1}{3}]
$$
By the following parametrization
\begin{equation}\label{parametr}
    \begin{array}{cc}
  |\alpha_{_{0}}|=\sin\theta \cos\varphi & |\beta_{_{0}}|=\sin\theta' \cos\varphi' \\
  |\alpha_{_{1}}|=\sin\theta \sin\varphi & |\beta_{_{1}}|=\sin\theta' \sin\varphi' \\
  \hspace{-.9cm}|\alpha_{_{2}}|=\cos\theta  & \hspace{-.9cm}|\beta_{_{2}}|=\cos\theta' \\
\end{array}
\end{equation}
where $\theta,\theta',\varphi,\varphi'\in[0,\frac{\pi}{2}]$, the
above statement can be written as
$$
3\cos\theta \cos\ \theta'+\frac{3}{2}[\sin^{2}\theta
\sin^{2}\theta' \cos
2(\varphi-\varphi')+\cos^{2}\theta+\cos^{2}\theta'-5\cos^{2}\theta
\cos^{2}\theta'-\frac{1}{3}]-1
$$
Since the Cauchy-Schwartz inequality becomes an equality when the
two vectors are parallel, we must have $\theta =\theta'$ and
$\varphi=\varphi'$. Then
$$
\mathrm{lhs}\leq 3\cos^{2}\theta -6\cos^{4}\theta.
$$
The lhs attains the maximum value $\frac{3}{8}$ when $\cos^{2}
\theta=\frac{1}{4}$.
\par
As implied from the above consideration, the pure product states
which saturate the maximum value $\frac{3}{8}$  are of the form
\begin{equation}\label{pure product}
    |\gamma\rangle=|\alpha\rangle\otimes|\beta\rangle=\frac{\sqrt{3}}{2}\left(%
\begin{array}{c}
  \cos\varphi \\
  e^{i\delta_{_{1}}}\sin\varphi \\
  \frac{1}{\sqrt{3}}\ e^{i\delta_{_{2}}} \\
\end{array}%
\right)\otimes\frac{\sqrt{3}}{2}\left(%
\begin{array}{c}
 \cos\varphi \\
  e^{i\delta_{_{1}}}\sin\varphi \\
  \frac{1}{\sqrt{3}}\ e^{i\delta_{_{2}}}\\
\end{array}%
\right)
\end{equation}
in which $\delta_{_{1}}$ and $\delta_{_{2}}$ are arbitrary real
phases.\\
{\bf The proof of (\ref{apboffplane1})}:\\
We have
$$
\frac{3}{4}(P_{_{1}}+P_{_{2}}+P_{_{1,2}}+P_{_{2,1}}+P_{_{3}}+P_{_{4}}+P_{_{5}}
 +P_{_{4,5}}+P_{_{5,4}}+P_{_{6}}+P_{_{7}}+P_{_{6,7}}+P_{_{7,6}}+P_{_{8}})-\frac{7}{4}
$$
$$
=3[\mathrm{Re}(\alpha_{_{0}}^{*}\alpha_{_{1}})\mathrm{Re}(\beta_{_{0}}^{*}\beta_{_{1}})
+\mathrm{Im}(\alpha_{_{0}}^{*}\alpha_{_{1}})\mathrm{Im}(\beta_{_{0}}^{*}\beta_{_{1}})
+\mathrm{Re}(\alpha_{_{0}}^{*}\alpha_{_{1}})\mathrm{Im}(\beta_{_{0}}^{*}\beta_{_{1}})
+\mathrm{Im}(\alpha_{_{0}}^{*}\alpha_{_{1}})\mathrm{Re}(\beta_{_{0}}^{*}\beta_{_{1}})
$$
$$
+\mathrm{Re}(\alpha_{_{0}}^{*}\alpha_{_{2}})\mathrm{Re}(\beta_{_{0}}^{*}\beta_{_{2}})
+\mathrm{Im}(\alpha_{_{0}}^{*}\alpha_{_{2}})\mathrm{Im}(\beta_{_{0}}^{*}\beta_{_{2}})
+\mathrm{Re}(\alpha_{_{0}}^{*}\alpha_{_{2}})\mathrm{Im}(\beta_{_{0}}^{*}\beta_{_{2}})
+\mathrm{Im}(\alpha_{_{0}}^{*}\alpha_{_{2}})\mathrm{Re}(\beta_{_{0}}^{*}\beta_{_{2}})
$$
$$
+\mathrm{Re}(\alpha_{_{1}}^{*}\alpha_{_{2}})\mathrm{Re}(\beta_{_{1}}^{*}\beta_{_{2}})
+\mathrm{Im}(\alpha_{_{1}}^{*}\alpha_{_{2}})\mathrm{Im}(\beta_{_{1}}^{*}\beta_{_{2}})
+\mathrm{Re}(\alpha_{_{1}}^{*}\alpha_{_{2}})\mathrm{Im}(\beta_{_{1}}^{*}\beta_{_{2}})
+\mathrm{Im}(\alpha_{_{1}}^{*}\alpha_{_{2}})\mathrm{Re}(\beta_{_{1}}^{*}\beta_{_{2}})]
$$
$$
\hspace{-4.4cm}+\frac{3}{4}(|\alpha_{_{0}}|^{2}-|\alpha_{_{1}}|^{2})(|\beta_{_{0}}|^{2}-|\beta_{_{1}}|^{2})
+\frac{1}{4}(1-3|\alpha_{_{2}}|^{2})(1-3|\beta_{_{2}}|^{2})-\frac{7}{4}
$$
$$
\hspace{-1cm}=3|\alpha_{_{0}}||\alpha_{_{1}}||\beta_{_{0}}||\beta_{_{1}}|
(\cos(\theta_{_{01}}-\theta'_{_{01}})+\sin(\theta_{_{01}}+\theta'_{_{01}}))
+3|\alpha_{_{0}}||\alpha_{_{2}}||\beta_{_{0}}||\beta_{_{2}}|
(\cos(\theta_{_{02}}-\theta'_{_{02}})
$$
$$
\hspace{0.5cm}+3|\alpha_{_{0}}||\alpha_{_{2}}||\beta_{_{0}}||\beta_{_{2}}|
(\cos(\theta_{_{02}}-\theta'_{_{02}})+\sin(\theta_{_{02}}+\theta'_{_{02}}))
+3|\alpha_{_{1}}||\alpha_{_{2}}||\beta_{_{1}}||\beta_{_{2}}|
(\cos(\theta_{_{12}}-\theta'_{_{12}})+\sin(\theta_{_{12}}+\theta'_{_{12}}))
$$
$$
\hspace{-4.5cm}+\frac{3}{4}(|\alpha_{_{0}}|^{2}-|\alpha_{_{1}}|^{2})(|\beta_{_{0}}|^{2}-|\beta_{_{1}}|^{2})
+\frac{1}{4}(1-3|\alpha_{_{2}}|^{2})(1-3|\beta_{_{2}}|^{2})-\frac{7}{4}
$$
$$
\hspace{-1cm}\leq
6|\alpha_{_{2}}||\alpha_{_{2}}|(|\alpha_{_{0}}||\beta_{_{ 0}}|
+|\alpha_{_{1}}||\beta_{_{ 1}}|+|\alpha_{_{2}}||\beta_{_{ 2}}|)
-\frac{15}{4}|\alpha_{_{2}}|^2|\beta_{_{2}}|^2+\frac{3}{4}(|\alpha_{_{0}}||\beta_{_{0}}|
+|\alpha_{_{1}}||\beta_{_{1}}|)^2
$$
$$
-\frac{3}{4}(|\alpha_{_{0}}||\beta_{_{1}}|
-|\alpha_{_{1}}||\beta_{_{0}}|)^2-\frac{3}{4}(|\alpha_{_{2}}|^2+|\beta_{_{2}}|^2)-\frac{3}{2}.
$$
By using the Cauchy-Schwartz inequality, we have
$$
\leq
6|\alpha_{_{2}}||\alpha_{_{2}}|-\frac{15}{4}|\alpha_{_{2}}|^2|\beta_{_{2}}|^2
+\frac{3}{4}[(|\alpha_{_{0}}||\beta_{_{0}}|+|\alpha_{_{1}}||\beta_{_{1}}|)^2-(|\alpha_{_{0}}||\beta_{_{1}}|
-|\alpha_{_{1}}||\beta_{_{0}}|)^2]-\frac{3}{4}(|\alpha_{_{2}}|^2+|\beta_{_{2}}|^2)-\frac{3}{2}.
$$
The parametrization (\ref{parametr}) yields
$$
=6\cos\theta\cos\theta'-\frac{15}{4}\cos^{2}\theta\
\cos^{2}\theta'
+\frac{3}{4}[\sin^{2}\theta\sin^{2}\theta'\cos2(\varphi-\varphi')-\cos^{2}\theta
-\cos^{2}\theta']-\frac{3}{2}.
$$
Since the Cauchy-Schwartz inequality becomes an equality when the
two vectors are parallel, we must have $\theta =\theta'$ and
$\varphi=\varphi'$. Then
$$
\leq 3\cos^{2}\theta -\cos^{4}\theta-\frac{3}{4}.
$$
The maximum value of the latter expression is 0 which come from
$\cos^{2}\theta=\frac{1}{2}$.

{\Large{Appendix III :}}\\
{\bf Proving the (\ref{witapp})}: \\
To prove (\ref{witapp}), we try to find all pure states that are
orthogonal to all pure product states (\ref{pure product}). To
this aim, let us consider a pure state as
$$
|\psi\rangle=\sum_{i,j=0}^{2}a_{_{ij}}|ij\rangle.
$$
and try to find the coefficients $a_{_{ij}}$ such that
$|\psi\rangle$ be orthogonal to all pure product states (\ref{pure
product}). We note that
\begin{equation}\label{pure}
    \begin{array}{c}
       \langle\gamma|\psi\rangle=a_{_{00}}(3\cos^{2}\varphi)+a_{_{01}}(3e^{-i\delta_{_{1}}}\sin\varphi\
\cos\varphi)+a_{_{02}}(\sqrt{3}e^{-i\delta_{_{2}}}\cos\varphi)+a_{_{10}}(3e^{-i\delta_{_{1}}}\sin\varphi\
\cos\varphi)\\
       +a_{_{11}}(3e^{-2i\delta_{_{1}}}\sin^{2}\varphi)+a_{_{12}}(\sqrt{3}e^{-i(\delta_{_{1}}
+\delta_{_{2}})}\sin\varphi)+a_{_{20}}(\sqrt{3}e^{-i\delta_{_{2}}}\cos\varphi)\\
    +a_{_{21}}(\sqrt{3}e^{-i(\delta_{_{1}}+\delta_{_{2}})}\sin\varphi)+a_{_{22}}e^{-i(\delta_{_{1}}+\delta_{_{2}})}.\\
   \end{array}
\end{equation}
For $\varphi=0$, we get
$$
3a_{_{00}}+\sqrt{3}e^{-i\delta_{_{2}}}(a_{_{02}}+a_{_{20}})+a_{_{22}}e^{-i(\delta_{_{1}}+\delta_{_{2}})}=0
$$
and arbitrariness of phases $\delta_{_{1}}$ and $\delta_{_{2}}$
follows that
$$
a_{_{00}}=a_{_{22}}=0\quad,\quad a_{_{20}}=-a_{_{20}}.
$$
$\varphi=\frac{\pi}{2}$ yields
$$
3a_{_{11}}e^{-2i\delta_{_{1}}}+\sqrt{3}e^{-i(\delta_{_{1}}+\delta_{_{2}})}(a_{_{12}}+a_{_{21}})=0
$$
and, as before, arbitrariness of phases $\delta_{_{1}}$ and
$\delta_{_{2}}$ follows that
$$
a_{_{11}}=0\quad,\quad a_{_{21}}=-a_{_{12}}.
$$
Substituting the above results in (\ref{pure}), we have
$a_{_{10}}=-a_{_{01}}$. Finally, it is concluded that the mutually
orthogonal pure states
\begin{equation}\label{psi}
    |\psi_{_{1}}\rangle=\frac{1}{\sqrt{2}}(|01\rangle-|10\rangle)\quad,\quad
|\psi_{_{2}}\rangle=\frac{1}{\sqrt{2}}(|02\rangle-|20\rangle)\quad,\quad
|\psi_{_{3}}\rangle=\frac{1}{\sqrt{2}}(|12\rangle-|21\rangle)
\end{equation}
are orthogonal to all pure product states (\ref{pure product}) and
the positive operator $\mathcal{P}$ in (\ref{witapp}) is
$$
\mathcal{P}=\frac{27}{4}|\psi_{_{1}}\rangle\langle\psi_{_{1}}|
+\frac{3}{4}(|\psi_{_{2}}\rangle\langle\psi_{_{2}}|+|\psi_{_{3}}\rangle\langle\psi_{_{3}}|).
$$
Now, we consider another pure state as
$$
|\varphi\rangle=\sum_{i,j=0}^{2}b_{_{ij}}|ij\rangle.
$$
and try to find the coefficients $b_{_{ij}}$ such that
$|\varphi\rangle$ be orthogonal to all pure product states
$|\alpha^{*}\rangle\otimes|\beta\rangle$ in (\ref{pure product}).
By similar arguments as above, we conclude that
$$
|\varphi\rangle=\frac{1}{\sqrt{11}}\
(|00\rangle+|11\rangle-3|22\rangle)
$$
and
$$
\mathcal{Q}=\frac{33}{8}\ |\varphi\rangle\langle\varphi|.
$$

\end{document}